\documentclass[sigconf]{acmart}
\renewcommand\footnotetextcopyrightpermission[1]{} 

\usepackage[utf8]{inputenc}

\usepackage[shortlabels]{enumitem}

\title[Is your chatbot GDPR compliant?]{Is your chatbot GDPR compliant? Open issues in agent design}

\acmConference[To appear at CUI'20: Conversational User Interfaces Conference]{CUI'20: International Conference on Conversational User Interfaces}{2020}{Spain}

\begin{document}

\author{Rahime Belen Sağlam}
\affiliation{University of Kent, UK}
\email{R.Belen-Saglam-724@kent.ac.uk}
\author{Jason R. C. Nurse}
\affiliation{University of Kent, UK}
\email{J.R.C.Nurse@kent.ac.uk}

\begin{abstract}
Conversational agents open the world to new opportunities for human interaction and ubiquitous engagement. As their conversational abilities and knowledge has improved, these agents have begun to have access to an increasing variety of personally identifiable information and intimate details on their user base. This access raises crucial questions in light of regulations as robust as the General Data Protection Regulation (GDPR). This paper explores some of these questions, with the aim of defining relevant open issues in conversational agent design. We hope that this work can provoke further research into building agents that are effective at user interaction, but also respectful of regulations and user privacy.
\end{abstract}


\maketitle

\section{Introduction}
Conversational agents are used in various contexts including the home, in finance, health care and tourism. In each context, personal information is increasingly collected and processed in order to provide more effective and personalized services. Personalization allows a chatbot to be aware of situations and to dynamically adapt its interaction to better suit user needs~\cite{neururer2018perceptions,chaves2019should}. Although deeper disclosures of personal information can increase the effective use of these agents, an intriguing---and thus far unaddressed---tension arises between the need for sharing such information (in one-off situations, or for personalized interactions) and privacy (a primary goal of regulations such as the GDPR). These points call attention to the question: is it possible to develop ultimately useful, GDPR-compliant (and thus privacy-aware) agents?

\section{Principles, lawful bases and rights}
The EU's GDPR is one of the most robust regulations for data protection that the world has ever seen. It defines principles and the lawful bases for processing  personal information, and also specifies rights for individuals~\cite{gdpr}. Below, we introduce a relevant subset of these which are then further explored in Section~\ref{open} to highlight key open issues in conversational agent practice and research.

\vspace{1em}
\noindent \textbf{Principles}:
\begin{itemize}
    \item \textit{Transparency}: requires data controllers to be clear, open and honest about how they process personal data.
    \item \textit{Data minimization}: requires data controllers ensure that personal data processed is adequate, relevant and limited to what is necessary in relation to the processing purpose.
    \item \textit{Purpose limitation}: requires personal data be collected for specified, explicit and legitimate purposes and not further processed.
    \item \textit{Storage limitation}: dictates that data controllers must delete personal data when it is no longer needed.
\end{itemize}

\noindent \textbf{Lawful basis for processing}:
\begin{itemize}
    \item \textit{Consent}: requires data controllers to obtain explicit consent from the data subject for the processing of  any  personal  data; this can be withdrawn at any time.
    \item \textit{Legitimate interest}: one of the cases where data controllers do not need to obtain consent is when they have a legitimate need and can show that the processing is necessary to achieve it.
    \item \textit{Special category data}: requires controllers to apply a higher level of protection for special categories of personal data (racial or ethnic origin, health data, political opinions etc.).
\end{itemize}

\noindent \textbf{Individual rights}:
\begin{itemize}
    \item\textit{Right to be informed}: allows individuals to know what is being done with their information, and thus links to transparency.
    \item \textit{Right of access}: allows data subjects to ask for a copy of their personal data, the purposes of processing their data, the categories of the data being processed, and the third parties or categories of third parties that will receive their data.
    \item \textit{Right to rectification}: requires data controllers to rectify or erase inaccurate or out of date information.
    \item \textit{Right to erasure}: also known as the right to be forgotten, mandates that controllers delete data in certain cases if there is no longer a lawful basis for processing or if the data subject withdraws consent.
\end{itemize}

\section{Open issues in agent design}
\label{open}

Even though GDPR and its implications have been widely covered in different contexts such as cloud computing, internet-of-things and blockchain technologies, it is surprising to see that very limited emphasis has been placed on potential design/implementation issues in the chatbot context. Even in those  that touch on GDPR compliance~\cite{peras2018chatbot,skjuve2018chatbots}, discussions are severely limited. Below, we explore key conflicts and open issues in conversational agent design. 

\subsection{How to build honest and open chatbots?}
Firstly, a lack of algorithmic \textit{transparency} is a major barrier for GDPR compliance in chatbots. Efforts towards making users more aware of how their personal information is processed are present but are rather constrained in scope~\cite{neururer2018perceptions,lai2018banking}. This limitation also becomes a challenge for data subjects when attempting to exercise their \textit{right to be informed}. Providing transparency becomes of the utmost importance for companies in the finance and health sectors, which provide personalised chatbots heavily processing sensitive or personally identifiable information. The issue here, therefore, is how is transparency best achieved in chatbot design, and how should users be kept informed about how their data is used?

Another relevant right, the \textit{right of access}, introduces key open issues since it is not clear how agents should/could provide access to the personal information they hold. Meeting this requirement is directly related to the accuracy of the agent in processing the conversations. Dissimilar to traditional applications that use relational databases, an agent has to process and extract personal information from a dialog. The risks are two-fold; the agent might not be able to extract the personal information or it may not process it accurately due to a failure in processing the text or voice. Those risks may undermine the trust between the agent and the user, and also make it very complicated for users to exercise their \textit{right to rectification}. Providing the entire conversation upon request to access personal data may be an option but it is not at all user friendly.

\subsection{How to design consent practices?}
How to manage \textit{consent} in chatbot applications gives rise to some other significant questions. One possible approach for gathering explicit consent for chatbots is assuming the activity of using a chatbot is innately giving consent. However, this will not meet the standard of an unambiguous indication by clear affirmative action in GDPR (see Article 4). It is possible to require users  to  `sign' a contract to obtain consent, or to gather it at the beginning of the conversation. In the latter option, it is difficult to judge the potentially negative impact on user-agent experience. Such kinds of formal and unusual treatment of language may fit in some use cases like finance applications well, however it may undermine the acceptance of a therapist chatbot, for instance. Accuracy of the agent to process the response of users could be another challenge. For example, if instead of giving the simple answer, ``yes, I consent'' or ``I do not'', they say ``ok, I consent but you cannot process my secrets about my family especially my husband!''. Then, the approach adopted by traditional applications (mobile apps, websites etc.), where individuals are asked to actively opt in to consent, might be an option for chatbots as well. This needs to be explored.

\subsection{Personalised chatbots vs. the right to be forgotten and the storage limitation principle }
Compliance is also challenged by difficulties in guaranteeing \textit{right to erasure}. Even though it could be technically straightforward to delete the previous conversations of a user, this will make personalisation impossible and undermine effective use of agents. For instance, Amazon Alexa allows deletion of voice recordings but also informs their users about potential problems: ``Voice recordings are used to improve the accuracy of your interactions with Alexa. Deleting voice recordings associated with your account may degrade your experience''~\cite{amaz}. A better solution could be deletion of personal data that seemed to be too sensitive to users after a dialog. However, questions then arise about how to request such a deletion from a chatbot and whether deletion should cover the entire two-way conversation. For example should agents support requests like, ``forget everything I told you after arguing with my boss'', ``forget my ethnic origin'' or ``don't store conversations we have about my mental health''? Anonymisation techniques could be applied as done in several domains to comply with GDPR, however, there are likely to be additional challenges in assuring anonymisation given the more fluid nature of conversational data. 

The \textit{storage limitation} principle could also be of concern for conversational agents. For a mobile application, it is reasonable to argue that the personal information of a user becomes useless when they deregister an application and there is no reason to keep it. However, in AI-based, intelligent applications like chatbots, one possible counterargument is that processing is necessary for the \textit{legitimate interests} of the data controller (e.g., a chatbot developer). The controller needs this information to train the agent which could be argued as a lawful basis to keep the data. However, this approach may not be inline with the \textit{purpose limitation} principle which prevents personal data to be processed further for a new purpose that is not compatible with the original one. Anonymisation by removing the identifiers of a person after a period of time may be the optimal solution for compliance. However, the exact meaning of `erasure' is  ambiguous for this solution. It is possible to argue that erasure refers to an outright deletion of entire conversations considering the possibility of identifying a data subject by other personal information they shared in the messages. How should this work, therefore?
 
\subsection{How to handle unneeded personal information?}
The interactive and conversational nature of chatbots also provides a challenge  for the \textit{data minimisation} principle. A chatbot may end up processing several items of sensitive personal information even if it does not expect (nor has it asked for) it. For instance, a user may prefer to disclose their ethnic origin while answering the questions of an agent about their stress level intentionally or unintentionally. Or, in a finance case, they may disclose account number and PIN to get an account balance (see~\cite{cnetamazon,guardiangoogle}). In such a case, in theory, it may be expected for an agent to avoid storing this information. Those inputs will inevitably be processed to generate an appropriate response. It is hard to find the correct strategy for a chatbot so that it can still give reasonable replies to the user and fully respect their privacy at the same time, especially in cases where \textit{special categories of personal data} (e.g., ethnicity or sexual orientation) are processed. It may be technically possible to avoid asking sensitive questions, however, we should keep in mind that the answer may still expose sensitive information. How, therefore, should agents be designed to cater for such eventualities?

In summary, while there is a need for personal user information to design and develop chatbots, it is also important to consider principles, lawful bases and rights under regulations such as the GDPR. This is clearly an area in need of more research as we, as a society, attempt to balance the advantages of agents with the need for privacy and the respective data protection laws and regulations. 
\bibliography{main}
\bibliographystyle{IEEEtran}
\end{document}